\documentclass[letterpaper, 10 pt, onecolumn]{ieeeconf}  

\usepackage{amsmath} 
\usepackage{amsfonts}
\usepackage{mathrsfs}
\usepackage{graphicx}
\usepackage{sgame}

\usepackage{soul} 
\usepackage{cite}
\usepackage{verbatim} 
\usepackage{svg}
\usepackage{float}
\usepackage{subcaption}

\graphicspath{{fig/}}

\newtheorem{theorem}{Theorem}[section]
\newtheorem{proposition}[theorem]{Proposition}
\newtheorem{example}[theorem]{Example}

\newtheorem{lemma}[theorem]{Lemma}

\title{\LARGE\bf
Information Design for Vehicle-to-Vehicle Communication
}

\author{Brendan T. Gould and Philip N. Brown
\thanks{A preliminary version of this work appeared in~\cite{gould_partial_2022}.}%
\thanks{Brendan T. Gould and Philip N. Brown are with the department of Computer Science at the University of Colorado Colorado Springs, Colorado Springs, CO, USA {\tt\small \{bgould2,pbrown2\}@uccs.edu}}%
\thanks{This work was supported in part by the Undergraduate Research Academy at the University of Colorado Colorado Springs and in part by NSF Award \#ECCS-2013779.}
}

\DeclareMathOperator*{\argmin}{arg\,min}

\newcommand{\xn}{x_{\rm n}}
\newcommand{\xv}{x_{\rm v}}
\newcommand{\xvu}{x_{\rm vu}}
\newcommand{\xvs}{x_{\rm vs}}

\newcommand{\xne}{x^{\rm ne}}
\newcommand{\xnne}{x_{\rm n}^{\rm ne}}
\newcommand{\xvune}{x_{\rm vu}^{\rm ne}}
\newcommand{\xvsne}{x_{\rm vs}^{\rm ne}}

\newcommand{\cn}{\chi_{\rm n}}
\newcommand{\cvu}{\chi_{\rm vu}}
\newcommand{\cvs}{\chi_{\rm vs}}

\newcommand{\Jn}{J_{\rm n}}
\newcommand{\Jvu}{J_{\rm vu}}
\newcommand{\Jvs}{J_{\rm vs}}

\newcommand{\Pn}{P_{\rm n}}
\newcommand{\Pvu}{P_{\rm vu}}
\newcommand{\Pvs}{P_{\rm vs}}

\newcommand{\A}{{\rm A}}
\renewcommand{\S}{{\rm S}} 

\newcommand{\C}{{\rm C}}
\newcommand{\R}{{\rm R}}

\begin{document}

\maketitle
\thispagestyle{empty}
\pagestyle{empty}

\begin{abstract}
The emerging technology of Vehicle-to-Vehicle (V2V) communication over vehicular \emph{ad hoc} networks promises to improve road safety by allowing vehicles to autonomously warn each other of road hazards.
However, research on other transportation information systems has shown that informing only a subset of drivers of road conditions may have a perverse effect of increasing congestion.
In the context of a simple (yet novel) model of V2V hazard information sharing, we ask whether partial adoption of this technology can similarly lead to undesirable outcomes.
In our model, drivers individually choose how recklessly to behave as a function of information received from other V2V-enabled cars, and the resulting aggregate behavior influences the likelihood of accidents (and thus the information propagated by the vehicular network).
We fully characterize the game-theoretic equilibria of this model using our new equilibrium concept.
Our model indicates that for a wide range of the parameter space, V2V information sharing surprisingly increases the equilibrium frequency of accidents relative to no V2V information sharing, and that it may increase equilibrium social cost as well.
\end{abstract}

\section{Introduction}
\label{sec:intro}
Technology is becoming increasingly intertwined with the society it serves, accelerated by emerging paradigms such as the internet of things (IoT) and various smart infrastructure concepts such as vehicle-to-vehicle communication (V2V).
It is no longer appropriate to design the merely-technical aspects of systems in isolation; rather, engineers must explicitly consider the implicit feedback loop between designed autonomy and human decision-making.
As a piece of this process, recent research has asked when new technological solutions may cause more harm than good~\cite{Brown2020a}.

A clear example of this is the area of equilibrium traffic congestion under selfish individual behavior. 
This topic has been well researched, and it is commonly understood that equilibria associated with this behavior may not be optimal at the system level~\cite{gairing_selfish_2008, correa_selfish_2004, dafermos_traffic_1984, wardrop_road_1952, massicot_public_2019, wu_information_2019}.
Many proposed solutions to this problem focus on the effects of deploying smart infrastructure to alleviate congestion and safety issues~\cite{nie_decentralized_2016}, using incentive design~\cite{Ferguson2021, lazar_learning_2021, biyik_incentivizing_2021} and information design~\cite{Zhu2020,massicot_competitive_2021} to improve upon selfish network routing.
However, this technology does not always have its intended effect; for example, self-driving cars can exacerbate equilibrium traffic congestion~\cite{Mehr2019}.

\emph{Bayesian persuasion} describes the process of a sender disclosing or obfuscating information in an attempt to influence the actions of other strategic agents~\cite{kamenica_bayesian_2011, bergemann_information_2019, akyol_information-theoretic_2017}.
However, it is known that merely making a subgroup of a population aware of a new road in a network can increase the equilibrium cost to that group, a phenomenon known as informational Braess' paradox~\cite{acemoglu_informational_2018, roman_how_2019}.
In information design problems in general, full disclosure of information is not always optimal~\cite{liu_effects_2016, massicot_public_2019, sayin_hierarchical_2019, tavafoghi_informational_2017,ukkusuri_information_2013,bergemann_information_2019}.

This naturally gives rise to the question of ``What is the optimal information sharing policy?''
Prior research has posed this question in the context of congestion games where each driver's cost depends on the selected route and the total mass of drivers on that route \cite{ dafermos_traffic_1984, lazar_learning_2021, wu_information_2019, ben-akiva_dynamic_1991}.

In this paper, we initiate a study on the incentive effects of distributed hazard information sharing by V2V-equipped vehicles, and ask when maximal sharing optimizes driver safety.
We pose a simple model of information sharing with partial V2V adoption; that is, some vehicles are unable to receive signals warning of road hazards.
In contrast to existing literature, our model allows for \emph{endogenous} road hazards where the likelihood of a road hazard is dependent on the aggregate recklessness of drivers.


After fully characterizing the emergent behavior in terms of a new equilibrium concept we call a \emph{signaling} equilibrium (Theorem~\ref{thm:EQCharacterization}), our main result in Theorem~\ref{thm:signalingCrashProbOptimization} shows that there exist parameter regimes in which the optimal hazard signaling rate is 0; that is, sharing any road hazard information with only V2V-equipped vehicles leads to a higher frequency of accidents than sharing none.
We then close with a discussion of the relationship between the \emph{social cost} and the frequency of accidents, and show that these two objectives are sometimes fundamentally opposed to one another: a signaling policy which decreases the frequency of accidents may necessarily increase the social cost  (and vice-versa).

\section{Model and Performance Metrics}
\label{section:model}

\subsection{Game Setup}
We adopt a nonatomic game formulation; i.e., we model a population of drivers as a continuum in which each of the infinitely-many drivers makes up an infinitesimally small portion of the population.
Each driver can choose to drive carefully (C), or recklessly (R), and a traffic accident either occurs ($\A$) or does not occur ($\neg \A$).
Reckless drivers may become involved in existing accidents and experience an expected accident cost of $r>1$; however, careful drivers regret their caution (e.g., due to the longer trip time incurred) if an accident is not present and experience a regret cost of $1$.
These costs are collected in this matrix:
\begin{center}
	\begin{game}{2}{2}
		\relax&Accident ($\A$)&No Accident ($\neg \A$)\\
		Careful ($\C$)&$0$&$1$\\
		Reckless ($\R$)&$r$&$0$\\
	\end{game}
\end{center}

We write $d\in[0,1]$ to denote the overall fraction of drivers choosing to drive recklessly, and $p(d)$ to represent the resulting probability that an accident occurs. 
Throughout the manuscript, we assume that more reckless drivers make an accident strictly more likely, so that $p(d)$ is strictly increasing.
Additionally, we assume that $p(d)$ is continuous. 

We model partial V2V adoption, i.e. some fraction $y~\in~[0, 1]$ of drivers have cars equipped with V2V technology.
When these drivers encounter road hazards or traffic accidents, the technology may autonomously detect these hazards and broadcast warning signals. 
If an accident has occurred, we say that V2V technology will detect the accident with probability $t(y)$. 
When an accident is detected, the technology will broadcast a signal that is received by all other V2V cars.
Furthermore, if no accident has occurred, we allow for the possibility that V2V technology incorrectly broadcasts a ``false-positive'' signal; this happens with probability $f(y)$. 
We assume that the technology broadcasts more true positives than false positives, i.e. $f(y) < t(y)$. 

In many models of transportation information systems, it is known that distributing perfect information can actually make parts or all of the population worse off~\cite{acemoglu_informational_2018, liu_effects_2016, sayin_hierarchical_2019, tavafoghi_informational_2017}; that is, the \emph{information design} problem is nontrivial: in some scenarios it may be optimal to withhold information from drivers.
Accordingly, we wish to study the information design problem faced by the administrators of V2V technology.
Therefore, let $\S$ be the event that a V2V car displays a warning to its driver, given that it has received a signal, and let $\beta := \mathbb{P}(\S) \in [0, 1]$.

This signaling scheme divides the population into three groups. 
We call a driver a \emph{non-V2V driver} if their vehicle lacks V2V technology, and a V2V driver otherwise.
We further differentiate V2V drivers by whether they have seen a warning signal, calling them \emph{unsignaled V2V drivers} if they have not seen a warning and \emph{signaled V2V drivers} if they have.
We write $\xn$, $\xvu$, and $\xvs$ to represent the mass of reckless drivers in each group, respectively, and a behavior profile as $x = (\xn, (\xvu, \xvs))$.

\begin{figure}[ht]
    \centering
    \includegraphics{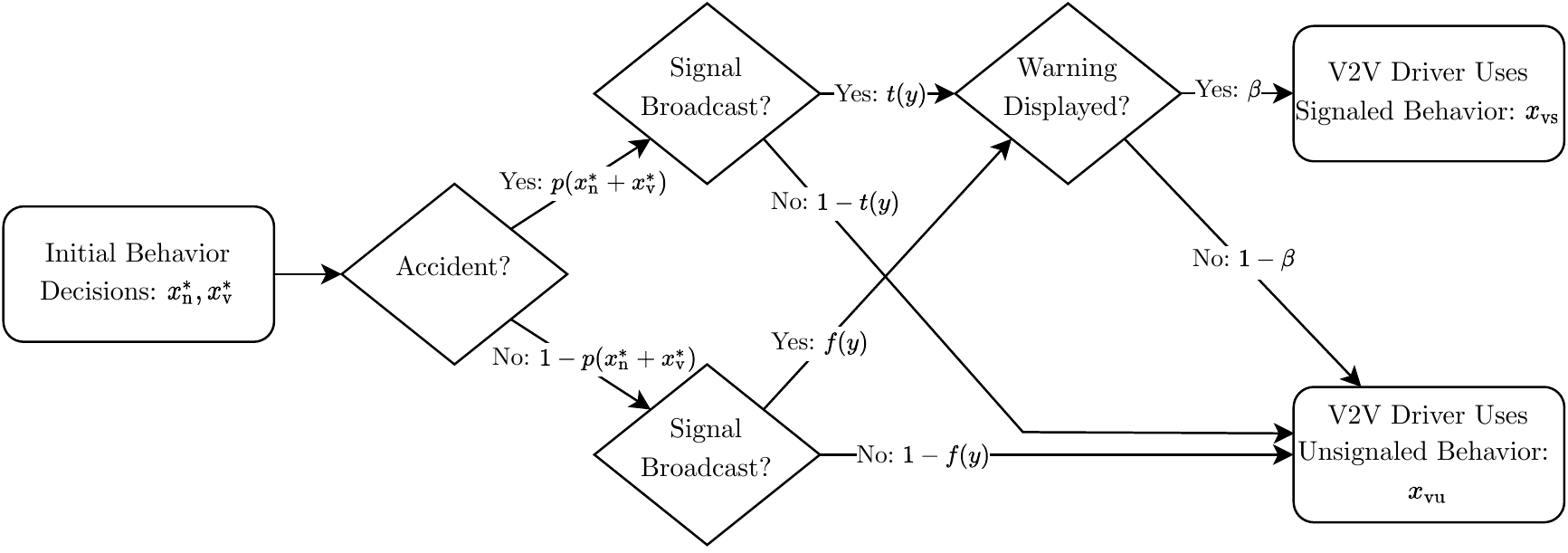}
    \caption{Timeline of behavior decisions}
    \label{fig:timeline}
\end{figure}

We assume that both non-V2V and V2V drivers have habitual behaviors $\xn^*$ and $\xv^*$, respectively.
Initially, each group of drivers makes their behavior decision according to their habits, and this behavior determines the probability of an accident. 
Non-V2V drivers receive no information that could lead them to change their behavior, and are thus effectively committed to their initial choice, i.e. $\xn = \xn^*$.
However, V2V drivers are able to adjust their behavior based on whether or not they see a warning signal; choosing unsignaled behavior $\xvu$ when they do not see a warning, and signaled behavior $\xvs$ when they do. 
The habitual behavior of V2V drivers must be a weighted average of their behavior when they do and do not see warnings, i.e.
\begin{equation}
    \xv^* = \mathbb{P}(\neg \S)\xvu + \mathbb{P}(\S)\xvs. \label{eq:avgBehavior}
\end{equation}
Figure \ref{fig:timeline} displays the overall timeline of events and decisions in our model. 

Define $P(x)$ to be the probability that an accident occurs, given some behavior profile $x$.
Additionally define $Q(x)$ as the probability that a given V2V driver sees a warning light, given the same. 
When the dependence on $x$ is clear from context, we will sometimes write simply $P$ and $Q$. 
Then, $\mathbb{P}(\A) = P(x) = p(\xn^* + \xv^*)$ and ${\mathbb{P}(\S) = Q(x) = \beta(P(x) t(y) + (1-P(x)) f(y))}$.
Note that $P(x)$ is implicitly parameterized by $\beta$ and $y$.
Substituting this into~\eqref{eq:avgBehavior} gives

\begin{equation}
    P(x) = p(\xn + (1-Q(x))\xvu + Q(x)\xvs ). \label{eq:consistency}
\end{equation}

\begin{proposition}
\label{prop:consistencySol}
For all parameter combinations, \eqref{eq:consistency} has at least one solution for $P(x)$. 
\end{proposition}
See Appendix~\ref{apx:consistencySolProof} for a proof of Proposition~\ref{prop:consistencySol}.


We write $\Jn (a;x)$ to denote the expected cost to a non-V2V driver choosing action $a \in \{\C, \R\}$, and similarly $\Jvu (a;x)$ and $\Jvs (a;x)$ for unsignaled and signaled V2V drivers, respectively. 
These costs are given by
\begin{align}
    \Jn (a;x)&=\begin{cases}
        1-P(x) & \text{if $a=\C$},\\
        rP(x) & \text{if $a=\R$},
        \label{eq:NV2VAverageCost}
    \end{cases} \\
    \Jvu (a;x)&=\begin{cases}
        1 - \mathbb{P}(\A | \neg \S) & \text{if $a=\C$},\\
        r \mathbb{P}(\A | \neg \S) & \text{if $a=\R$},
        \label{eq:V2VUnsignaledCost}
    \end{cases} \\
    \Jvs (a;x)&=\begin{cases}
        1 - \mathbb{P}(\A | \S) & \text{if $a=\C$},\\
        r \mathbb{P}(\A | \S) & \text{if $a=\R$}.
        \label{eq:V2VSignaledCost}
    \end{cases}
\end{align}
Finally, we define a signaling game as the tuple $G~=~(\beta, y, r)$.

\subsection{Signaling Equilibrium}
We define a signaling equilibrium as a behavior profile $\xne~=~(\xnne, (\xvune, \xvsne))$ with $0 \le \xnne, \xvsne, \xvune \le 1$ satisfying the following:
\begin{align}
    \xnne < 1-y &\implies \Jn (\C; \xne) \leq \Jn (\R; \xne), \label{eq:NV2VCarefulInUse} \\
    \xnne > 0 &\implies \Jn (\R; \xne) \leq \Jn (\C; \xne), \label{eq:NV2VRecklessInUse} \\
    \xvune < y &\implies \Jvu (\C; \xne) \leq \Jvu (\R; \xne), \label{eq:V2VUnsignaledCarefulInUse} \\
    \xvune > 0 &\implies \Jvu (\R; \xne) \leq \Jvu (\C; \xne), \label{eq:V2VUnsignaledRecklessInUse} \\
    \xvsne < y &\implies \Jvs (\C; \xne) \leq \Jvs (\R; \xne), \label{eq:V2VSignaledCarefulInUse} \\
    \xvsne > 0 &\implies \Jvs (\R; \xne) \leq \Jvs (\C; \xne). \label{eq:V2VSignaledRecklessInUse}
\end{align}
Equations \eqref{eq:NV2VCarefulInUse}-\eqref{eq:V2VSignaledRecklessInUse} enforce the standard conditions of a Nash equilibrium (i.e. if players are choosing any action, its cost to them is minimal).
The novelty of this concept lies in the fact that we endogenously determine the likelihood of a signal, and therefore the mass of signaled and unsignaled V2V drivers, using accident probability at equilibrium.
But accident probability is determined by driver behavior, which is in turn influenced by the probability of a signal. 
This creates a complex interdependence between driver behavior and signal probability, which is captured by our consistency equation \eqref{eq:consistency}. 

Additionally, we define social cost as the expected individual cost given behavior: 
\begin{equation}
\begin{split}
    S(x) &= \Jn (\C; x) (1 - y - \xn) + \Jn (\R; x) \xn \\
    &+ (1 - Q) (\Jvu (\C; x) (y - \xvu) + \Jvu (\R; x) \xvu) \\
    &+ Q(\Jvs (\C; x) (y - \xvs) + \Jvs (\R; x) \xvs).
\end{split}
\label{eq:socialCost}
\end{equation}

\begin{proposition}
For every signaling game $G$, there exists a signaling equilibrium $\xne$ and it is essentially unique.
By this we mean that for any two signaling equilibria $\xne_1$ and $\xne_2$ of $G$, both of the following hold:
\begin{align}
    \xne_{\rm{n}1} + (1-Q(\xne_1))\xne_{\rm{vu}1} + Q(\xne_1)\xne_{\rm{vs}1} &= \xne_{\rm{n}2} + (1-Q(\xne_2))\xne_{\rm{vu}2} + Q(\xne_2)\xne_{\rm{vs}2}, \label{eq:uniqueMass} \\
    P(\xne_1) &= P(\xne_2). \label{eq:uniqueCrashProb}
\end{align}
\end{proposition}
We provide a proof of this fact in Lemma \ref{lemma:EQTypes}.
A tedious series of arguments can show that \eqref{eq:uniqueMass} and \eqref{eq:uniqueCrashProb} imply $S(\xne_1) = S(\xne_2)$, further supporting the notion that these equilibria are effectively the same. 
Throughout this paper, we use the terms ``unique'' and ``essentially unique'' interchangeably to mean \eqref{eq:uniqueMass} and \eqref{eq:uniqueCrashProb} are satisfied.%
\footnote{This definition is motivated by the degenerate cases where $\beta = 0$. 
If this is the case, then $Q=0$, meaning V2V cars will never receive signals or display warnings to their drivers. 
Therefore, there is no real distinction between non-V2V drivers and V2V drivers, which causes many different behavior tuples to be effectively identical.}

\subsection{Research Objectives}
Our first objective is to characterize the signaling equilibria of any game $G$.
In Theorem~\ref{thm:EQCharacterization}, we show that every game $G$ has an essentially unique signaling equilibrium. 
Additionally, we show that receiving a signal makes V2V drivers more cautious and not receiving a signal makes them more reckless at equilibrium, compared to non-V2V drivers.

Next, we seek to optimize accident probability and social cost by means of signal display probability.
To that end, we abuse notation and write $P(G)$ to denote $P(\xne)$ and $S(G)$ to denote $S(\xne)$ where $\xne$ is a signaling equilibrium of game $G$.
We then wish to find values for $\beta_P$ and $\beta_S$ such that
\begin{gather}
    \beta_P \in \argmin_{\beta \in [0, 1]}{P(G)}, \label{eq:signalingCrashProbOptimization} \\
    \beta_S \in \argmin_{\beta \in [0, 1]}{S(G)}. \label{eq:signalingSocialCostOptimization}
\end{gather}

\begin{figure}
    \centering
    \vspace{2mm}
    \includegraphics[scale=0.6]{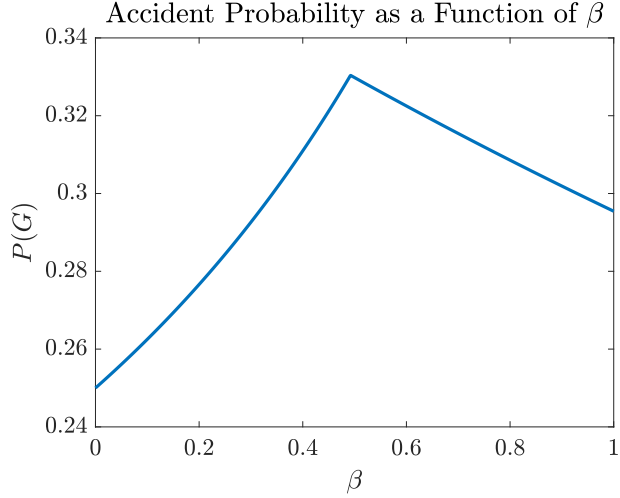}
    \caption{Equilibrium accident frequency with respect to the signal display probability $\beta$.
    Note that when $\beta<0.5$, displaying more warning signals steeply increases the frequency of accidents, and that even the maximum possible signal display probability $\beta=1$ yields a higher accident frequency than if warning signals were never shown to drivers.
    The example depicted has accident probability characterized by $p(d)=0.3d+0.1$, signal probability characterized by $t(y)=0.8y$ and $f(y)=0.1y$, V2V penetration $y=0.9$, and accident cost $r=3$.}
    \label{fig:PofG}
    \vspace{-3mm}
\end{figure}

In Theorem~\ref{thm:signalingCrashProbOptimization}, we provide an algorithm to determine a solution to \eqref{eq:signalingCrashProbOptimization}, and show that there exist games where $\beta_P = 0$ is a solution, as depicted in Figure~\ref{fig:PofG}.
Furthermore, in Proposition~\ref{prop:signalingSocialCostOptimization}, we provide sufficient criteria for when $\beta_S = 1$ is a solution to \eqref{eq:signalingSocialCostOptimization}, and show that there paradoxically exist regions of the parameter space where $\beta_S = 1$ is \emph{not} a solution to \eqref{eq:signalingSocialCostOptimization}.
Furthermore, each of these paradoxes can still occur if $f(y) \equiv 0$, indicating that poor quality technology is not their sole cause. 
See Appendix~\ref{apx:noFPCounterExamples} for a list of examples of the paradoxes in this case. 

\section{Our Contributions}
\subsection{Equilibrium Characterization}
A signaling equilibrium takes the form of a tuple listing the mass of reckless drivers in each of our three population groups.
These masses implicitly determine an equilibrium crash probability through \eqref{eq:consistency}.
Though this relationship is complicated, our first theorem shows that an equilibrium is uniquely determined by any given parameter combination. 

\begin{theorem}
\label{thm:EQCharacterization}
For any V2V signaling game $G=(\beta,y,r)$, a signaling equilibrium exists and is essentially unique.
In particular, the equilibrium $\xne = (\xnne, (\xvune, \xvsne))$  can take one of the following 4 forms:
\begin{itemize}
    \item $(0, (0, 0))$,
    \item $(0, (\cvu, 0))$, for some $\cvu \in [0, y]$,
    \item $(\cn, (y, 0))$, for some $\cn \in [0, 1-y]$,
    \item $(1-y, (y, \cvs))$, for some $\cvs \in [0, y]$.
\end{itemize}
\vspace{2mm}
\end{theorem}

Theorem~\ref{thm:EQCharacterization} captures several important characteristics of signaling equilibria. 
Chief among these is the fact that a signaling equilibrium exists and is essentially unique for every game $G = (\beta, y, r)$.
Additionally, note that there is an ``ordering'' to recklessness: every unsignaled V2V driver must be reckless before any non-V2V driver can be, and all non-V2V drivers must be reckless before any signaled V2V driver can. 
This is because V2V technology has made signaled V2V drivers more confident that an accident has occurred, and therefore more likely to drive carefully than non-V2V drivers. 
Similarly, unsignaled V2V drivers have an extra measure of confidence that an accident has \emph{not} occurred, and are therefore more likely to drive recklessly than non-V2V drivers.

The proof of Theorem~\ref{thm:EQCharacterization} appears at length in Section~\ref{sec:proofs}.
However, we provide here a key necessary condition of any signaling equilibrium; it serves as a cornerstone of the proof and will be instrumental in later results:


\begin{lemma}
\label{lemma:EQConditions}
For any signaling game $G = (\beta, y, r)$, a behavior profile $\xne = (\xnne, (\xvune, \xvsne))$ is a signaling equilibrium if the following hold:
\begin{align}
    \xnne &= \begin{cases}
        0 & \text{if } P(\xne) > \frac{1}{1+r}, \\
        p^{-1}(P(\xne)) - (1 - Q(\xne))y & \text{if } P(\xne) = \frac{1}{1+r}, \\
        1-y & \text{if } P(\xne) < \frac{1}{1+r},
        \label{eq:NV2VChoice}
    \end{cases} \\
    \xvune &= \begin{cases}
        0 & \text{if } \mathbb{P} (\A | \neg \S)  > \frac{1}{1+r}, \\
        \frac{p^{-1}(P(\xne))}{1-Q(\xne)} & \text{if } \mathbb{P} (\A | \neg \S)  = \frac{1}{1+r}, \\
        y & \text{if } \mathbb{P} (\A | \neg \S)  < \frac{1}{1+r}.
        \label{eq:V2VUnsignaledChoice}
    \end{cases} \\
    \xvsne &= \begin{cases}
        0 & \text{if } \mathbb{P} (\A | \S)  > \frac{1}{1+r}, \\
        \frac{p^{-1}(P(\xne)) - 1 + Q(\xne)y}{Q(\xne)} & \text{if } \mathbb{P} (\A | \S)  = \frac{1}{1+r}, \\
        y & \text{if } \mathbb{P} (\A | \S)  < \frac{1}{1+r}.
        \label{eq:V2VSignaledChoice}
    \end{cases}
\end{align}
Furthermore, this equilibrium is essentially unique (i.e all signaling equilibria that exist satisfy \eqref{eq:uniqueMass} and \eqref{eq:uniqueCrashProb}). 
\vspace{2mm}
\end{lemma}

The remainder of this section contains a brief overview of techniques and notation that provide useful understanding of the problem; a full proof of Lemma \ref{lemma:EQConditions} and Theorem \ref{thm:EQCharacterization} is contained in Section \ref{sec:proofs}.

It can be shown using Bayes' Theorem that
\begin{align}
    \def\arraystretch{0.5}
    \label{eq:V2VUnsignaledBayes}
    \mathbb{P} (\A | \neg \S ) 
    \begin{array}{c}
    <  \\ = \\ >
    \end{array}
    \frac{1}{1+r} \hspace{-0.8mm} &\iff \hspace{-0.8mm} P(x)
    \def\arraystretch{0.5}
    \begin{array}{c}
    < \\ = \\ >
    \end{array}
    \frac{1-\beta f(y)}{1 + r(1-\beta t(y)) -\beta f(y)}, \\
    \label{eq:V2VSignaledBayes}
    \mathbb{P} (\A | \S ) 
    \def\arraystretch{0.5}
    \begin{array}{c}
    <  \\ = \\ >
    \end{array}
    \frac{1}{1+r} \hspace{-0.8mm} &\iff \hspace{-0.8mm} P(x)
    \def\arraystretch{0.5}
    \begin{array}{c}
    < \\ = \\ >
    \end{array}
     \frac{f(y)}{rt(y)+f(y)}.
     \def\arraystretch{1.0}
\end{align}

We use this notation to mean that any of the relationships between the first expressions is equivalent to the corresponding relationship between the second expressions. 
Equality and both inequalities are preserved. 

Each of the above expressions acts as a threshold on the behavior of a particular group of drivers. 
For example, if $P(x) < \frac{1-\beta f(y)}{1 + r(1-\beta t(y)) -\beta f(y)}$, then by \eqref{eq:V2VUnsignaledBayes}, $\mathbb{P}(\A|\neg \S) < \frac{1}{1+r}$. 
Then, by Lemma \ref{lemma:EQConditions}, $\xvune = y$, meaning all unsignaled V2V cars are reckless. 
Similarly, if $P(x) > \frac{1-\beta f(y)}{1 + r(1-\beta t(y)) -\beta f(y)}$, \eqref{eq:V2VUnsignaledBayes} gives that $\mathbb{P}(\A|\neg \S) > \frac{1}{1+r}$, so Lemma \ref{lemma:EQConditions} guarantees that all unsignaled V2V cars are careful. 

In the same way, the expression $\frac{f(y)}{rt(y)+f(y)}$ acts as a threshold on the behavior of signaled V2V drivers, and $\frac{1}{1+r}$ does for non-V2V drivers. 
For convenience, we use the following shorthand to reference these thresholds:
\begin{equation}
    \Pvs := \frac{f(y)}{rt(y)+f(y)}, \hspace{3mm} \Pn := \frac{1}{1+r}, \hspace{3mm} \Pvu := \frac{1-\beta f(y)}{1 + r(1-\beta t(y)) -\beta f(y)},
\end{equation}
where it holds that:
\begin{equation}
    \label{eq:recklesnessOrdering}
    \Pvs < \Pn \leq \Pvu.
\end{equation}

Intuitively, this ordering corresponds to the fact that each group of drivers has different information about the world. 
Unsignaled drivers are more confident that an accident has not occurred (because they receive a signal some of the time that accidents do occur), and are therefore willing to ``risk'' driving recklessly at higher inherent accident probabilities than non-V2V drivers are. 
The opposite is true for signaled V2V drivers. 

\begin{figure}
    \centering
    \begin{subfigure}{0.49\textwidth}
        \includegraphics[width=\linewidth]{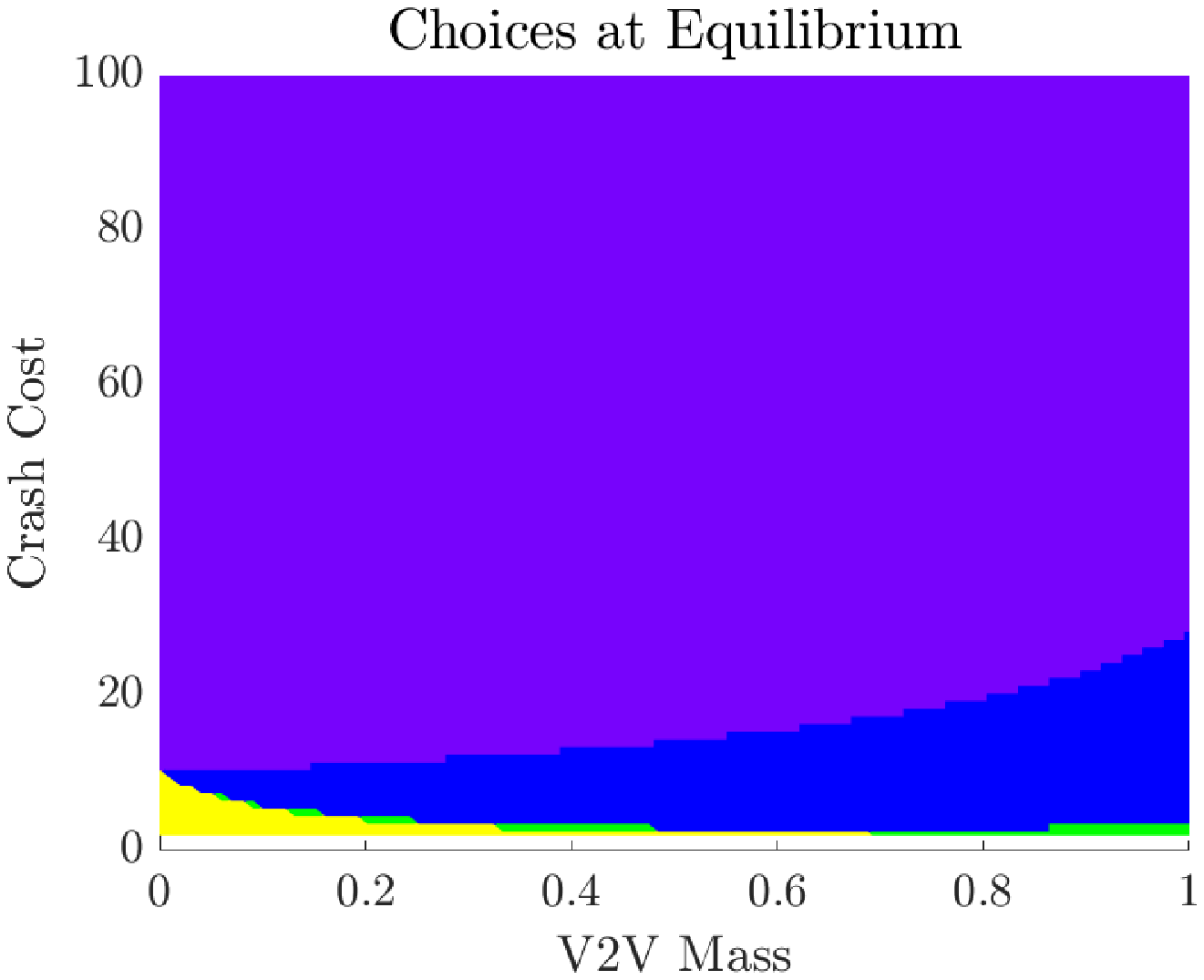}
        \caption{Parameterized by: $p(d) = 0.8d + 0.1$, $t(y) = 0.7y$, $f(y) = 0.1y$.}
        \label{subfig:EQPlot_Low_Reckless}
    \end{subfigure}
    \hfill
    \begin{subfigure}{0.49\textwidth}
        \includegraphics[width=\linewidth]{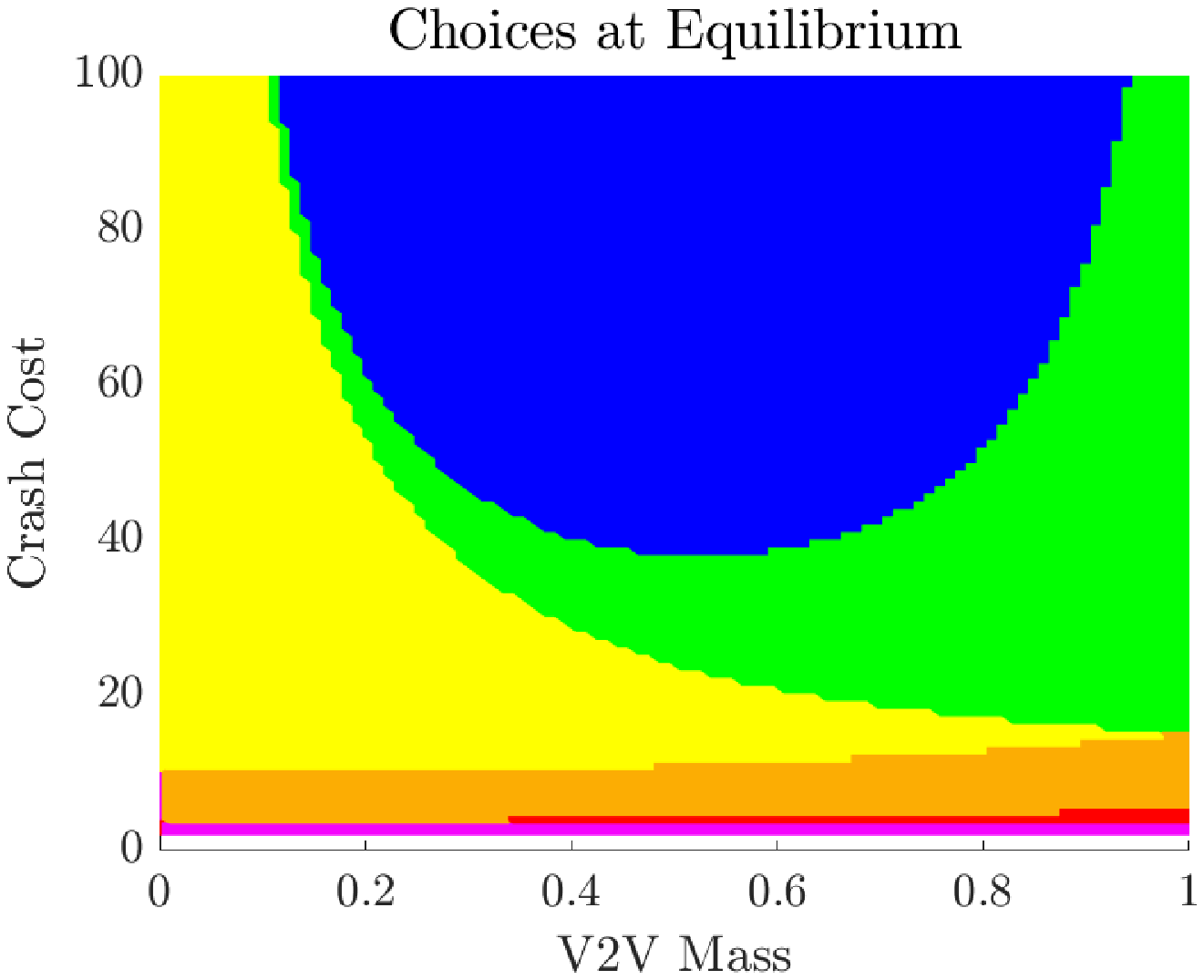}
        \caption{Parameterized by: $p(d) = 0.1d$, $t(y) = 0.95y$, $f(y) = 0.3y$.}
        \label{subfig:EQPlot_Mid_Reckless}
    \end{subfigure}
    \hfill
    \begin{subfigure}{0.49\textwidth}
        \includegraphics[width=\linewidth]{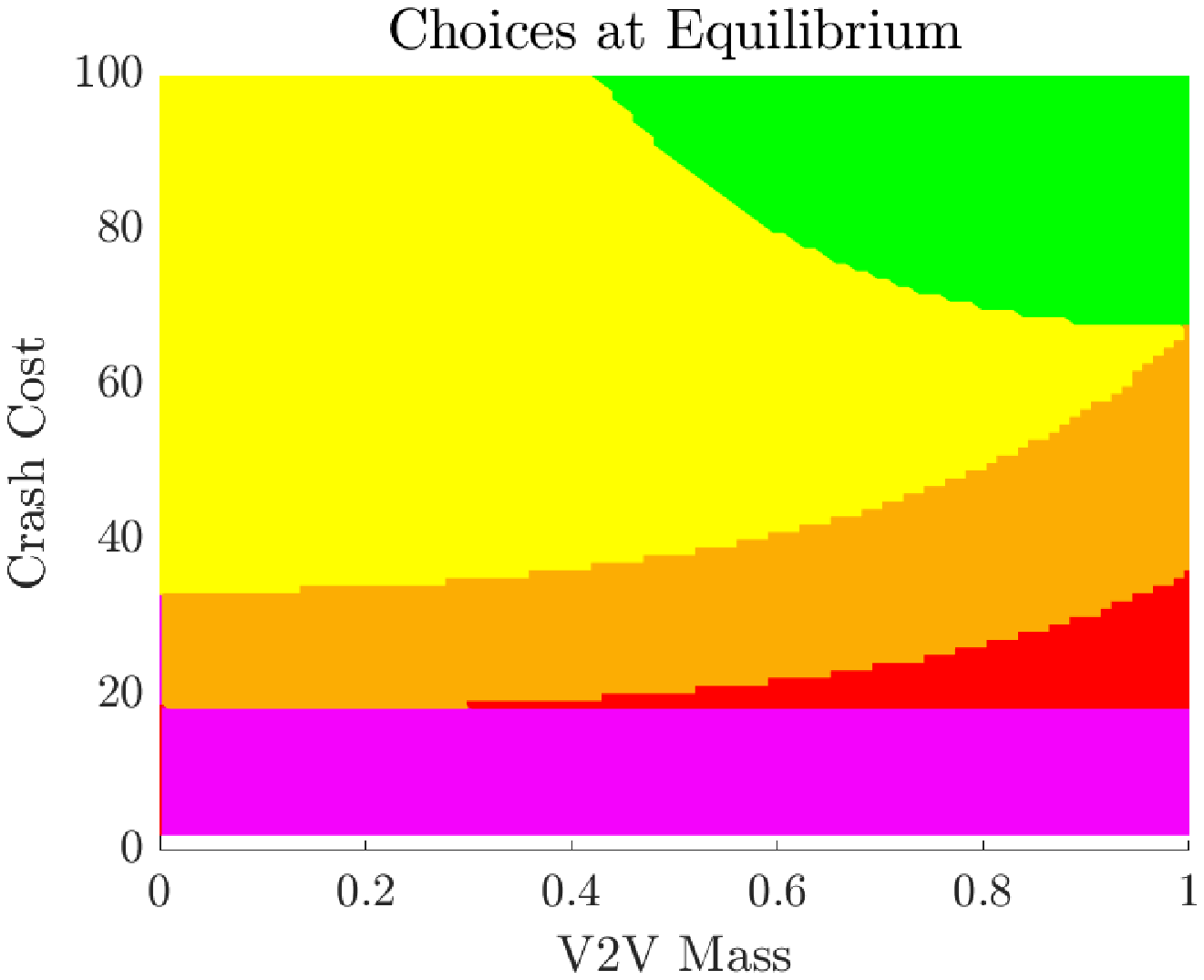}
        \caption{Parameterized by: $p(d) = 0.03d$, $t(y) = 0.95y$, $f(y) = 0.5y$.}
        \label{subfig:EQPlot_Hi_Reckless}
    \end{subfigure}
    \hfill
    \begin{subfigure}{0.49\textwidth}
        \centering
        \includegraphics[width=\linewidth]{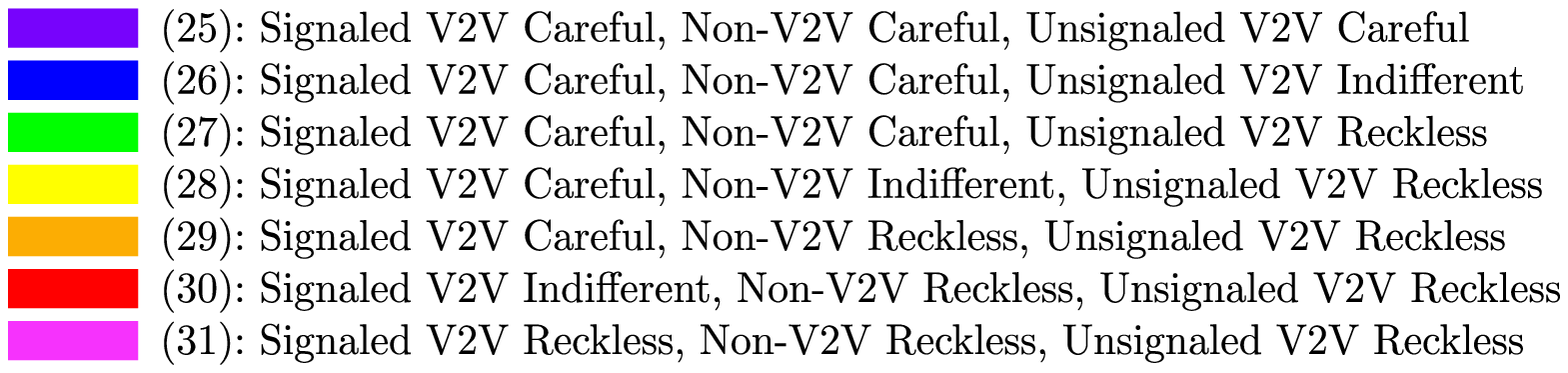}
        \label{subfig:EQPlot_Legend}
    \end{subfigure}

    \caption{Equilibrium ranges. 
    With $\beta=1$ fixed, these plots show the equilibrium range that is valid for a combination of the parameters $y$ and $r$. 
    Each colored range corresponds to one of $E_1$ to $E_7$. }
    \label{fig:EQ_Ranges}
\end{figure}

Based on these thresholds, we can determine the essentially unique equilibrium behavior tuple for any accident probability $P$. 
The sets below correspond to regions of parameter space that constrain the equilibrium accident probability with respect to our behavior thresholds (Lemma \ref{lemma:crashProbRanges}), and therefore allow us to compute behavior directly from the model parameters (Lemma \ref{lemma:EQTypes}). 
\begin{align}
    E_1 &:= \{(\beta, y, r) : \Pvu < p(0)\} \label{range:SCNCUC} \\
    E_2 &:= \{(\beta, y, r) : p(0) \le \Pvu \le  p((1 - \beta \Pvu (t(y)-f(y)) - \beta f(y))y) \label{range:SCNCUI}\} \\
    E_3 &:= \{(\beta, y, r) : p((1 - \beta \Pvu (t(y)-f(y)) - \beta f(y))y) < \Pvu \land \Pn < p((1 - \beta \Pn (t(y)-f(y)) - \beta f(y))y)\} \label{range:SCNCUR} \\
    E_4 &:= \{(\beta, y, r) : p((1 - \beta \Pn (t(y)-f(y)) - \beta f(y))y) \le \Pn \le p(1 - (\beta \Pn (t(y)-f(y)) + \beta f(y))y)\} \label{range:SCNIUR} \\
    E_5 &:= \{(\beta, y, r) : p(1 - (\beta \Pn (t(y)-f(y)) + \beta f(y))y) < \Pn \land \Pvs < p(1 - (\beta \Pvs (t(y)-f(y)) + \beta f(y))y)\} \label{range:SCNRUR} \\
    E_6 &:= \{(\beta, y, r) : p(1 - (\beta \Pvs (t(y)-f(y)) + \beta f(y))y) \le \Pvs \le p(1)\} \label{range:SINRUR} \\
    E_7 &:= \{(\beta, y, r) : p(1) < \Pvs\} \label{range:SRNRUR}
\end{align}
These sets are visualized for particular slices of parameter space in Figure \ref{fig:EQ_Ranges}. 

Finally, the equilibrium behavior within each range is consistent with the form claimed in Theorem \ref{thm:EQCharacterization}. 
This completes our conceptual overview of the characterization result; again, see Section~\ref{sec:proofs} for the complete proof. 

\subsection{Information Design for Minimizing Accident Probability}
\label{subsec:signalingOptimization}
Though common intuition would suggest that increasing the quality of information given to drivers would allow them to make more informed decisions and arrive at less costly outcomes, prior research has shown that this is not always the case 
\cite{acemoglu_informational_2018, roman_how_2019}. 
Because of this, it is a non-trivial question for V2V administrators to decide the optimal quality of information to distribute. 
This quality will be certainly be bound by technical limitations, but administrators can freely manipulate it within the feasible range. 

When a car with V2V technology receives a warning signal, it does not necessarily have to display a warning light to its driver. 
Paradoxically, we show that ignoring accidents in this way can decrease accident probability at equilibrium.

\begin{theorem}
\label{thm:signalingCrashProbOptimization}
For any signaling game $G = (\beta, y, r)$, we must have that either:
\begin{equation}
    0 \in \argmin_{\beta \in [0, 1]}{P(G)} \text{ or } 1 \in \argmin_{\beta \in [0, 1]}{P(G)}. \label{eq:signalingMinCrashProb} \\
\end{equation}
Furthermore, there exist signaling games where $\beta=1$ does not minimize accident probability.
\vspace{2mm}
\end{theorem}

In other words, the minimum accident probability is guaranteed to be caused by never displaying warnings, or by displaying them as often as technologically possible.
The proof of Theorem~\ref{thm:signalingCrashProbOptimization} proceeds as follows:
\begin{itemize}
\item First, Lemma~\ref{lemma:eqRangeOrdering} shows that the feasible equilibrium ranges in any game must satisfy a particular ordering with respect to $\beta$.
\item Next, Lemma~\ref{lemma:signalingCrashProbMonotonicity} uses this ordering to show that the probability of an accident is weakly increasing for low values of $\beta$, and weakly decreasing otherwise. 
Then, the smallest possible value of $\beta$ will always be a minimum within the increasing range, and the largest value of $\beta$ must be a minimum in the decreasing range. 
\item Therefore, one of the two must be a global minimum, completing the proof of Theorem~\ref{thm:signalingCrashProbOptimization}.
\end{itemize}

\begin{lemma}
\label{lemma:eqRangeOrdering}
For any combination of the parameters $y$ and $r$, let $\beta_1, \beta_2 \in [0, 1]$ and $\beta_1 < \beta_2$. 
Additionally, let $G_1 = (\beta_1, y, r)$ and $G_2 = (\beta_2, y, r)$.
Then the equilibrium type that is valid for $G_1$ and $G_2$ is ordered by $\beta$. 
Specifically, for any integer $i \in [1, 7]$, 
\begin{equation}
    G_1 \in E_i \implies G_2 \in E_j
\end{equation}
for some integer $j \in [i, 7]$. 
\end{lemma}

\begin{proof}
Let $P_{\rm vu1} = \frac{1-\beta_1 f(y)}{1 + r(1-\beta_1 t(y)) -\beta_1 f(y)}$ and $P_{\rm vu2}=\frac{1-\beta_2 f(y)}{1 + r(1-\beta_2 t(y)) -\beta_2 f(y)}$, and note that $P_{\rm vu1} \le P_{\rm vu2}$. 


If $G_1 \in E_2$, then $G_2 \not \in E_1$.
If it was, then we would have 
\[
P_{\rm vu2} < p(0) \le P_{\rm vu1},
\]
which is a clear contradiction.
Therefore, by Theorem \ref{thm:EQCharacterization}, $G_2 \in \bigcup_{i=2}^7 E_i$, which is the desired result. 

Similarly, if $G_1 \in E_3$, then $G_2 \not \in E_1 \cup E_2$.
If $G_2 \in E_1$, we would have that 
\[
P_{\rm vu2} < p(0) \le p((1 - \beta_1 P_{\rm vu1} (t(y)-f(y)) - \beta_1 f(y))y) < P_{\rm vu1},
\]
and if $G_2 \in E_2$, 
\[
P_{\rm vu2} \le p((1 - \beta_2 P_{\rm vu2} (t(y)-f(y)) - \beta_2 f(y))y) \le p((1 - \beta_1 P_{\rm vu1} (t(y)-f(y)) - \beta_1 f(y))y) < P_{\rm vu1}.
\]
(The second inequality holds since $\beta_1 < \beta_2$, $\Pvu$ is increasing in $\beta$, and $p$ is an increasing function.)
In any case, we have a contradiction. 
Again by Theorem \ref{thm:EQCharacterization}, $G_2 \in \bigcup_{i=3}^7 E_i$, completing the proof in this case. 

This technique can be extended using \eqref{eq:recklesnessOrdering} to cover the cases where $G_1 \in E_4$, $G_1 \in E_5$, $G_1 \in E_6$, or $G_1 \in E_7$. 
In any case, it is a matter of assuming $G_2$ is not contained in the claimed sets, finding a string of inequalities that leads to a contradiction, and applying Theorem \ref{thm:EQCharacterization} to achieve the desired result. 
\end{proof}

We will now provide a full definition of $P(G)$, the function optimized by \eqref{eq:signalingCrashProbOptimization}.
It is always true that $P(G)=P(\xne)$, but using the ranges defined by \eqref{range:SCNCUC}-\eqref{range:SRNRUR}, we can be more specific. 
For any signaling game $G = (\beta, y, r)$, the probability of an accident at its unique signaling equilibrium $\xne$ is given by
\begin{equation}
    P(G) = \begin{cases}
        p(0) & \text{if } G \in E_1, \\
        \Pvu & \text{if } G \in E_2, \\
        P(\xne) & \text{if } G \in E_3, \\
        \Pn & \text{if } G \in E_4, \\
        P(\xne) & \text{if } G \in E_5, \\
        \Pvs & \text{if } G \in E_6, \\
        p(1) & \text{if } G \in E_7.
    \end{cases}
    \label{eq:crashProbs}
\end{equation}
This simplification is due to Lemma~\ref{lemma:crashProbRanges}.
Since there exists a signaling equilibrium for any game $G$, this function is defined for all parameter combinations.
We now present a piecewise monotonicity result on $P(G)$ with respect to $\beta$.

\begin{lemma}
\label{lemma:signalingCrashProbMonotonicity}
There exists a $\bar{\beta}$ such that for $\beta \le \bar{\beta}$, $P((\beta, y, r))$ is weakly increasing in $\beta$, and for $\bar{\beta} < \beta$, $P((\beta, y, r))$ is weakly decreasing in $\beta$.
Unfortunately, the problem does not permit a closed form expression where this $\bar{\beta}$ is isolated; however, $\bar{\beta}$ is the quantity that satisfies
\[
    \frac{1-\bar{\beta} f(y)}{1 + r(1-\bar{\beta} t(y)) - \bar{\beta} f(y)} = p\left(\left(1 - \bar{\beta} \frac{1-\bar{\beta} f(y)}{1 + r(1-\bar{\beta} t(y)) - \bar{\beta} f(y)}(t(y)-f(y))-\bar{\beta} f(y) \right)y\right).
\]
\vspace{2mm}
\end{lemma}

\begin{proof}
For any combination of the parameters $y$ and $r$, let $\beta_1, \beta_2 \in [0, 1]$ and $\beta_1 < \beta_2$. 
Let $G_1 = (\beta_1, y, r)$ and $G_2 = (\beta_2, y, r)$.
Our approach is a exhaustive comparison of crash probabilities within and between the cases of \eqref{eq:crashProbs}. 
If
\[
    \frac{1-\beta_2 f(y)}{1 + r(1-\beta_2 t(y)) -\beta_2 f(y)} \leq p\left(\left(1 - \beta_2 \frac{1-\beta_2 f(y)}{1 + r(1-\beta_2 t(y)) -\beta_2 f(y)}(t(y)-f(y))-\beta_2 f(y) \right)y\right), 
\]
then accident probability is weakly increasing. 
(Note that this condition is merely a special case of the rightmost inequality described in \eqref{range:SCNCUI}.) 

If $p(0) \le \frac{1-\beta_2 f(y)}{1 + r(1-\beta_2 t(y)) -\beta_2 f(y)}$, $G_2 \in E_2$, and otherwise $G_2 \in E_1$. 
By Lemma \ref{lemma:eqRangeOrdering}, if $G_2\in E_2$, then either $G_1 \in E_1$ or $G_1 \in E_2$.
In either case, by \eqref{eq:crashProbs}, $P(G_1) \le P(G_2)$.
Otherwise, $G_2 \in E_1$, so $G_1 \in E_1$ as well (again by Lemma \ref{lemma:eqRangeOrdering}). 
Clearly, $p(0) \le p(0)$, so in any case, we have that $P(G_1) \le P(G_2)$, the desired result.

Now, consider sufficiently large values of $\beta$, i.e. assume
\[
    \frac{1-\beta_1 f(y)}{1 + r(1-\beta_1 t(y)) -\beta_1 f(y)} > p\left(\left(1 - \beta_1 \frac{1-\beta_1 f(y)}{1 + r(1-\beta_1 t(y)) -\beta_1 f(y)}(t(y)-f(y))-\beta_1 f(y) \right)y\right).
\]
We shall show that accident probability is weakly decreasing. 
(This condition is simply a special case of the leftmost inequality in \eqref{range:SCNCUR}.)

Note that $G_1 \not \in E_1$ and $G_1 \not \in E_2$, so $G_1 \in \bigcup_{i=3}^7 E_i$ by Theorem \ref{thm:EQCharacterization}. 
Very similar techniques to the above suffice in every case except when $G_1, G_2 \in E_3$ or $G_1, G_2 \in E_5$.  

If $G_1 \in E_3$ and $G_2 \in E_3$, then Lemma \ref{lemma:EQTypes} guarantees that $(0, (y, 0))$ is an equilibrium for both games. 
Then, by \eqref{eq:consistency},
\[
    P(x) = p((1-Q(x))y) = p((1 - \beta P(x)(t(y)-f(y))-\beta f(y)) y).
\]

Let $P_1$ and $P_2$ be the quantities that satisfy ${P_1 = p((1 - \beta P_1(t(y)-f(y))-\beta f(y)) y)}$ and $P_2 = p((1 - \beta P_2(t(y)-f(y))-\beta f(y)) y)$, respectively. 
By \eqref{eq:consistency}, $P(G_1) = P_1$ and $P(G_2) = P_2$.
Assume by way of contradiction that $P_1 \leq P_2$. 
We use algebraic manipulations to work ``up'' one level of recursion, starting with the definition of $\beta_1$ and $\beta_2$.
This gives that 
\[
    (1 - \beta_1 (P_1(t(y)-f(y))+f(y)))y > (1 - \beta_2 (P_2(t(y)-f(y))+f(y)))y.
\]
Since $p(d)$ is increasing, it preserves the inequality, so
\[
    p((1 - \beta_1 P_1(t(y)-f(y)) - \beta_1 f(y)))y) > p((1 - \beta_2 P_2(t(y)-f(y)) - \beta_2 f(y)))y).
\]
But then by definition of $P_1$ and $P_2$, we can substitute to obtain $P_1 > P_2$, contradicting our hypothesis. 
Therefore, we must have that $P(G_1) = P_1 > P_2 = P(G_2)$, the desired conclusion. 
If $G_1 \in E_5$ and $G_2 \in E_5$, a very similar technique can be used. 
Thus, $P$ is decreasing with $\beta$. 
\end{proof}

This result immediately gives a minimizing signal display probability in each range. 
We use this result to prove Theorem \ref{thm:signalingCrashProbOptimization}.

\subsubsection*{Proof of Theorem~\ref{thm:signalingCrashProbOptimization}}
Immediately from Lemma \ref{lemma:signalingCrashProbMonotonicity}, we have that either the smallest or largest value of $\beta$ must minimize $P$.
Therefore, either $0 \in \argmin_{\beta \in [0, 1]}{P(G)}$, or $1 \in \argmin_{\beta \in [0, 1]}{P(G)}$, as desired.

It remains to show that there actually exist signaling games such that $\beta_P = 1$ does not minimize accident probability. 
To that end, let $p(d) = 0.8d + 0.1$, $t(y) = 0.8y$, $f(y)=0.1y$, $y = 0.9$, and $r = 20$.
Then $G_0 = (0, y, r) \in E_1$, and $G_1 = (1, y, r) \in E_2$. 
Therefore, by \eqref{eq:crashProbs}, $P(G_0) = p(0) = 0.1$, and $P(G_1) = \frac{1-(1) f(y)}{1 + r(1-(1) t(y)) -(1) f(y)} \approx 0.1398$. 
Since $P(G_0) < P(G_1)$, $\beta_P = 1$ cannot be a solution to \eqref{eq:signalingCrashProbOptimization}.
\hfill\QED

\subsection{Information Design for Minimizing Social Cost}
\label{subsec:signalingOptimizationSocialCost}

It is also useful to consider how to minimize social cost at equilibrium. 
Again, intuition suggests that the social cost minimizing value of $\beta$ should be $1$, but this is not always the case.
We present the counter-intuitive result that there exist games where full information sharing among V2V drivers does not optimize social cost.

\begin{figure}
    \centering
    \begin{subfigure}{0.49\textwidth}
        \includegraphics[width=\linewidth]{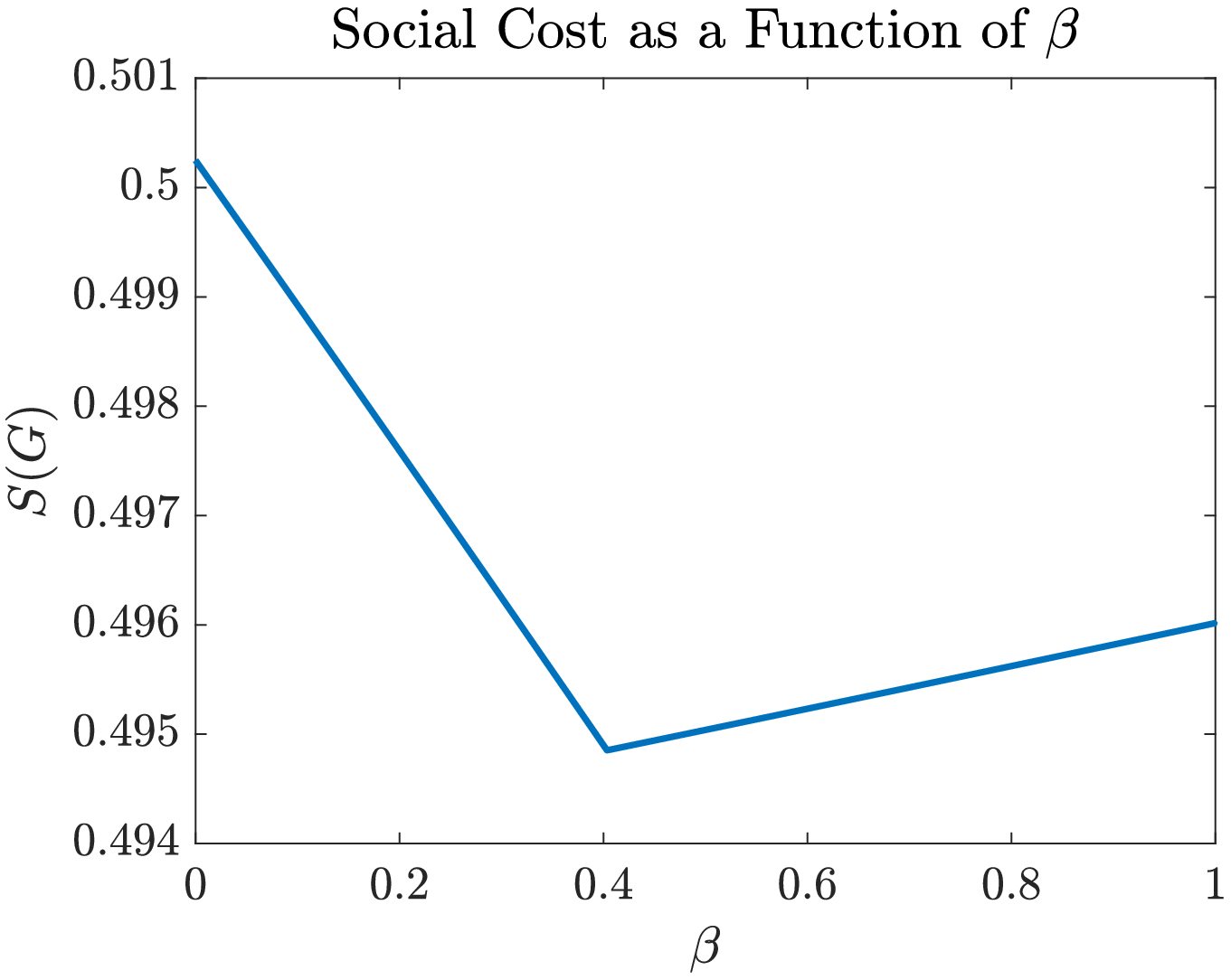}
        \caption{Social cost in $E_3$. 
        Accident probability characterized by $p(d) = d^\frac{1}{4}$, signal probability characterized by $t(y)=0.9y$ and $f(y)=0.1y$, V2V penetration $y=0.066$, and accident cost $r=1.001$.}
        \label{subfig:SofG_SCNCUR}
    \end{subfigure}
    \hfill
    \begin{subfigure}{0.49\textwidth}
        \includegraphics[width=\linewidth]{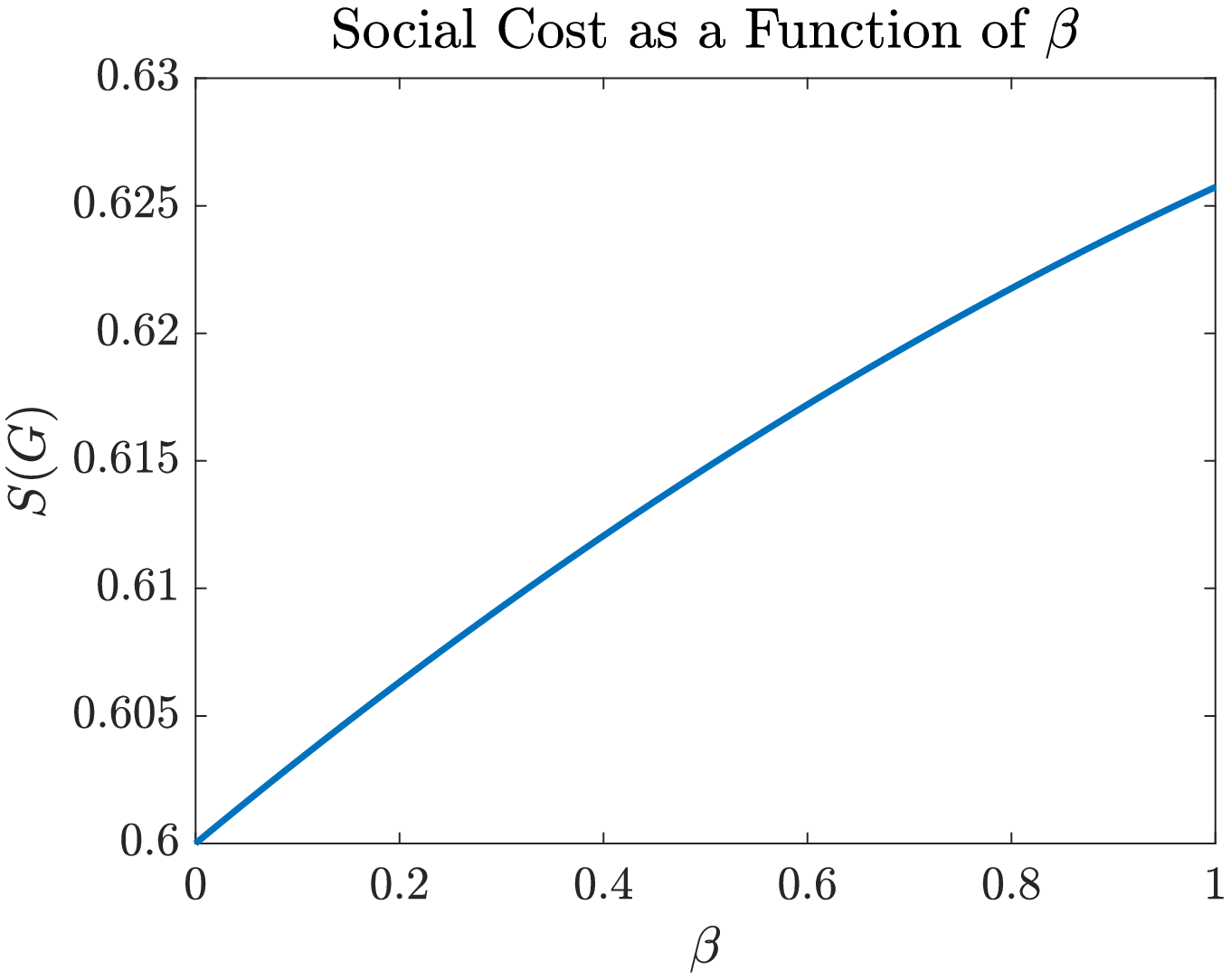}
        \caption{Social cost in $E_5$. 
        Accident probability characterized by $p(d) = 0.03d$, signal probability characterized by $t(y)=0.95y$ and $f(y)=0.5y$, V2V penetration $y=0.4$, and accident cost $r=20$.}
        \label{subfig:SofG_SCNRUR}
    \end{subfigure}
    \caption{Equilibrium social cost with respect to the signal display probability $\beta$. 
        Note the paradoxical result that there exist parameters where displaying more warning signals increases the social cost at equilibrium.}
    \label{fig:SofG}
\end{figure}

We provide examples to illustrate that $\beta_S$ need not be 1.
\begin{example}
\label{ex:SCNCURSocialCostIncreasing}
Let $p(d) = d^\frac{1}{4}$, $t(y) = 0.9y$, $f(y) = 0.1y$, $y = 0.066$, and $r = 1.001$. 
Additionally, let $\beta_1 = 0.4204$, $\beta_2 = 1$, $G_1 = (\beta_1, y, r)$, and $G_2 = (\beta_2, y, r)$.
Then, $G_1 \in E_3$ and $G_2 \in E_3$.
Numerical solvers give that $S(G_1) \approx 0.4949$, while $S(G_2) \approx 0.4960$.
Thus, $S(G_2) > S(G_1)$. 
This parameter set is visualized in Figure~\ref{subfig:SofG_SCNCUR}. 
\end{example}

\begin{example}
\label{ex:SCNRURSocialCostIncreasing}
Let $p(d) = 0.03d$, $t(y) = 0.95y$, $f(y) = 0.5y$, $y = 0.4$, and $r = 20$. 
Additionally, let $\beta_1 = 0$, $\beta_2 = 1$, $G_1 = (\beta_1, y, r)$, and $G_2 = (\beta_2, y, r)$.
Then, $G_1 \in E_5$ and $G_2 \in E_5$.
Numerical solvers give that $S(G_1) \approx 0.6$, while $S(G_2) \approx 0.6257$.
Thus, $S(G_2) > S(G_1)$. 
This parameter set is visualized in Figure~\ref{subfig:SofG_SCNRUR}.
\end{example}

In general, $S(G)$ is decreasing with respect to $\beta$, with two exceptions. 
Within $E_3$ and $E_5$, social cost may sometimes be increasing. 

\begin{proposition}
\label{prop:signalingSocialCostOptimization}
For any signaling game $G = (\beta, y, r)$, $S(G)$ is decreasing with respect to $\beta$ unless $G \in E_3$ or $G \in E_5$.
There exist games in both $E_3$ and $E_5$ where $S(G)$ is increasing with respect to $\beta$. 
\end{proposition}

\begin{proof}
First, consider the case where $G \in E_2$. 
By Lemma \ref{lemma:crashProbRanges}, $P(G) = \Pvu$. 
Furthermore, by Lemma \ref{lemma:EQTypes}, $\xne = \left(0, \left(\frac{p^{-1}(\Pvu)}{1-Q}, 0\right)\right)$ is the essentially unique signaling equilibrium for $G$. 
Then, \eqref{eq:socialCost} simplifies to 
\[
S(G) = r\frac{1-\beta t(y)}{1+r(1 - \beta t(y)) - \beta f(y)},
\]
which is decreasing in $\beta$. 
(This can be seen by simply taking the partial derivative with respect to $\beta$, and noting that it is negative.)
A similar technique suffices for the same result in the cases where $G \in E_1$, $G \in E_4$, $G \in E_6$, or $G \in E_7$. 

Now, consider the case where $G \in E_3$ or $G \in E_5$. 
Example \ref{ex:SCNCURSocialCostIncreasing} describes a game in $E_3$ where $S(G)$ is increasing with $\beta$.
Similarly, Example \ref{ex:SCNRURSocialCostIncreasing} describes a game in $E_5$ where $S(G)$ is increasing with $\beta$.
Thus, social cost need not be decreasing with $\beta$ in these ranges. 
\end{proof}

\begin{figure}
    \centering
    \includegraphics[scale=0.6]{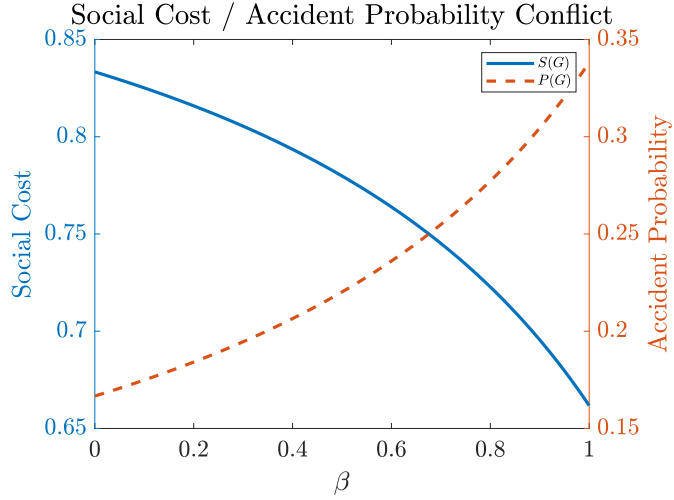}
    \caption{Equilibrium accident probability and social cost with respect to the signal display probability $\beta$. 
    Note that the social cost minimizing value of $\beta$ is 1, while the accident probability minimizing value of $\beta$ is 0. 
    The example depicted has accident probability characterized by $p(d) = 0.8d + 0.1$, signal probability characterized by $t(y)=0.9y$ and $f(y)=0.1y$, V2V penetration $y=0.8$, and accident cost $r=5$.}
    \label{fig:PSConflict}
\end{figure}

Note that if $G \in E_3$, $S(G)$ can be increasing with respect to $\beta$, but $P(G)$ is guaranteed to be decreasing by Lemma \ref{lemma:signalingCrashProbMonotonicity}.
Additionally, if $G \in E_2$, then $P(G)$ is increasing and $S(G)$ is decreasing. 
This implies that V2V administrators face an inherent trade-off in their optimization decision. 
To minimize accident probability, they must sometimes accept a higher than optimal social cost, and vice versa.
This conflict is depicted in Figure~\ref{fig:PSConflict}.

\section{Proofs of Theorem~\ref{thm:EQCharacterization}} \label{sec:proofs}
\subsubsection*{Proof of Lemma \ref{lemma:EQConditions}}
Assume that a behavior tuple $x = (\xn, (\xvu, \xvs))$ satisfies equations \eqref{eq:NV2VChoice}-\eqref{eq:V2VSignaledChoice}.
We shall show that $x$ is a signaling equilibrium. 

First, assume that $P(x) > \frac{1}{1+r}$. 
By basic algebra, $1-P < rP$, or equivalently by \eqref{eq:NV2VChoice}, $\Jn(\C; x) < \Jn(\R; x)$.
Then, simply note that $\xn = 0$ satisfies \eqref{eq:NV2VCarefulInUse} and \eqref{eq:NV2VRecklessInUse}.
Similarly, if $P = \frac{1}{1+r}$ we have that $\Jn(\C; x) = \Jn(\R; x)$, and if $P < \frac{1}{1+r}$, then $\Jn(\C; x) > \Jn(\R; x)$. 
In any case, \eqref{eq:NV2VChoice} forces $\xn$ to satisfy \eqref{eq:NV2VCarefulInUse} and \eqref{eq:NV2VRecklessInUse}.
An identical method shows that the conditions enforced by \eqref{eq:V2VUnsignaledChoice} satisfy \eqref{eq:V2VUnsignaledCarefulInUse} and \eqref{eq:V2VUnsignaledRecklessInUse}, and that \eqref{eq:V2VSignaledChoice} satisfies \eqref{eq:V2VSignaledCarefulInUse} and \eqref{eq:V2VSignaledRecklessInUse}.
Since \eqref{eq:NV2VCarefulInUse}-\eqref{eq:V2VSignaledRecklessInUse} are all satisfied, $x$ must be a signaling equilibrium. 

We will now show that any signaling equilibrium $x' = (\xn', (\xvu', \xvs'))$ is essentially identical to $x$. 
Consider \eqref{eq:NV2VChoice} in the case where $P(x') > \frac{1}{1+r}$. 
Assume by way of contradiction that $\xn' \ne \xn$, which immediately forces that $\xn' > 0$. 
Since $x'$ is a signaling equilibrium, by \eqref{eq:NV2VRecklessInUse}, we have that 
\[
\Jn(\R; x') \le \Jn(\C; x').
\]
But we showed above that because $P(x') > \frac{1}{1+r}$, 
\[
\Jn(\C; x') < \Jn(\R; x'),
\]
a clear contradiction. 
Therefore, $\xn' = \xn$. 
A very similar technique using \eqref{eq:NV2VCarefulInUse} shows that if $P(x') < \frac{1}{1+r}$, then $\xn' = 1-y = \xn$.
This technique can be further reused together with \eqref{eq:V2VUnsignaledCarefulInUse}-\eqref{eq:V2VSignaledRecklessInUse} to show that $\xvu' = \xvu$ in the first and third cases of \eqref{eq:V2VUnsignaledChoice} and $\xvs' = \xvs$ in the first and third cases of \eqref{eq:V2VSignaledChoice}. 
In any of these cases, we then have that $x' = x$, so clearly $x'$ is essentially identical to $x$.

It remains to show that $x'$ is essentially identical to $x$ in the second cases of \eqref{eq:NV2VChoice}-\eqref{eq:V2VSignaledChoice}. 
If $\mathbb{P}(\A | \S) = \frac{1}{1+r}$, then \eqref{eq:V2VSignaledBayes} and \eqref{eq:recklesnessOrdering} imply that $P(x') = \Pvs < \Pn \le \Pvu$. 
But then by \eqref{eq:V2VUnsignaledBayes}, $P(x') < \frac{1}{1+r}$ and $\mathbb{P}(\A|\neg\S) < \frac{1}{1+r}$. 
Therefore, by the above, $\xn' = 1-y$ and $\xvu' = y$, so by \eqref{eq:consistency},
\[
P(x') = p (1-y+y-Qy+Q\xvs) = p(1-Q(y-\xvs)).
\]
But then simple algebra gives that $\xvs' = \frac{p^{-1}(P(x')) - 1 + Q(x')y}{Q(x')}$, which proves \eqref{eq:V2VSignaledChoice}. 

If $\beta t(y) > 0$, then the inequality described in \eqref{eq:recklesnessOrdering} becomes strict, and a very similar technique suffices to show \eqref{eq:NV2VChoice} and \eqref{eq:V2VUnsignaledChoice}. 
Otherwise, we must have that $\beta t(y) = 0$. 
In the second case of either \eqref{eq:NV2VChoice} or \eqref{eq:V2VUnsignaledChoice}, \eqref{eq:consistency} simplifies to
\[
    P(x') = p(\xn' + \xvu') = \frac{1}{1+r},
\]
which implies that $p^{-1}\left(\frac{1}{1+r}\right) = \xn' + (1-Q)\xvu' + Q\xvs'$.
Since the same must be true for $x$, \eqref{eq:uniqueMass} and \eqref{eq:uniqueCrashProb} are satisfied, so the equilibria are essentially identical.
Therefore, the tuple satisfying \eqref{eq:NV2VChoice}-\eqref{eq:V2VSignaledChoice} is a signaling equilibrium, and is essentially unique. 
\hfill \QED

Recall that \eqref{range:SCNCUC}-\eqref{range:SRNRUR} divide our parameter space into seven regions. 
Lemma~\ref{lemma:crashProbRanges} shows that each of these regions restricts the possible values of $P$.

\begin{lemma}
\label{lemma:crashProbRanges}
For any signaling game $G = (\beta, y, r)$, 
\begin{equation}
   G \in \bigcup_{i=1}^7 E_i,  
\end{equation}
and each of these ranges restricts the possible values of $P(G)$ according to the following table. 
\vspace{2mm}
\begin{center}
\begin{tabular}{|c|c|c|c|c|c|c|}
    \hline
    $G \in E_1$ & $G \in E_2$ & $G \in E_3$ & $G \in E_4$ & $G \in E_5$ & $G \in E_6$ & $G \in E_7$ \\
    \hline 
    $P(G) = p(0)$ & $P(G) = \Pvu$ & $\Pn < P(G) < \Pvu$ & $P(G) = \Pn$ & $\Pvs < P(G) < \Pn$ & $P(G) = \Pvs$ & $P(G) = p(1)$ \\ 
    \hline
\end{tabular}
\end{center}
\vspace{2mm}

\end{lemma}

Each of these claims is proved via contradiction. 
Applying Lemma \ref{lemma:EQConditions} to the contradiction hypothesis places restrictions on the values of $\xnne$, $\xvune$, and $\xvsne$.
Next, using these values and \eqref{eq:consistency}, we perform algebraic operations to take $P$ ``up'' one level of its recursive definition. 
Finally, we show that this new expression for $P$ forces a contradiction. 

\begin{proof}
Consider any game $G = (\beta, y, r)$. 
If $G \not \in E_1$ and $G \not \in E_2$, then we must have that 
\[
p((1 - \beta \Pvu (t(y)-f(y)) - \beta f(y))y) < \Pvu.
\]
Similarly, if $G \not \in E_7$ and $G \not \in E_6$, then we must have that 
\[
p((1 - \beta \Pvu (t(y)-f(y)) - \beta f(y))y) < \Pvu.
\]
Therefore, if $G \not \in E_3$ and $G \not \in E_5$,  
\[
p((1 - \beta \Pn (t(y)-f(y)) - \beta f(y))y) \le \Pn \le p(1 - (\beta \Pn (t(y)-f(y)) + \beta f(y))y).
\]
But this implies that $G \in E_4$. 
Therefore, $G \in \bigcup_{i=1}^7 E_i$, as desired. 

Since the remaining technique is sufficient for all seven claims, we will prove only the case where $G \in E_3$, i.e.
\[
    p((1 - \beta \Pvu (t(y)-f(y)) - \beta f(y))y) < \Pvu \text{ and } \Pn < p((1 - \beta \Pn (t(y)-f(y)) - \beta f(y))y).
\]
Note that if $\beta t(y) = 0$, $\Pn = \Pvu$, and the above conditions simplify to $p(y) < \Pvu$ and $\Pn < p(y)$, respectively. 
But this is clearly a contradiction, since it implies that $\Pn < p(y) < \Pvu = \Pn$. 
Therefore, $\beta t(y) > 0$, and both inequalities described in \eqref{eq:recklesnessOrdering} are strict (i.e. $\Pn < \Pvu$). 

If we assume by way of contradiction that $P(\xne) \le \frac{1}{1+r}$, by \eqref{eq:V2VUnsignaledBayes} and because $\Pn < \Pvu$, $\mathbb{P} (\A | \neg \S) < \frac{1}{1+r}$. 
Therefore, by Lemma \ref{lemma:EQConditions}, $\xvune = y$.
Furthermore, it is always true that $\xnne \ge 0$ and $\xvsne \ge 0$. 
Then, starting with our contradiction hypothesis, we perform algebraic operations to take $P$ ``up'' one level of its recursive definition in \eqref{eq:consistency}. 
This gives that 
\[
\xnne +\xvune - \left(\beta P(\xne) (t(y)-f(y)) + \beta f(y) \right)(\xvune-\xvsne) \ge 0 + y - \left(\frac{\beta (t(y)-f(y))}{1+r} + \beta f(y)\right)(y-0).
\]

Since $p(d)$ is strictly increasing, it preserves the inequality, giving
\[
    p(\xnne +\xvune - \left(\beta P(\xne) (t(y)-f(y)) + \beta f(y) \right)(\xvune-\xvsne)) = P(\xne) \ge p((1 - \beta \Pn (t(y)-f(y)) - \beta f(y))y).
\]
Therefore, applying \eqref{range:SCNCUR}, we have that 
\[
    p((1 - \beta \Pn (t(y)-f(y)) - \beta f(y))y) \le P(\xne) \le \Pn < p((1 - \beta \Pn (t(y)-f(y)) - \beta f(y))y),
\]
an obvious contradiction. 
Therefore, we must have that $\frac{1}{1+r} = \Pn < P(\xne)$. 
This technique can also be used to show that $P(\xne) < \Pvu = \frac{1-\beta f(y)}{1+r(1-\beta t(y)) -\beta f(y)}$, completing the proof in this case. 

A proof of the remaining claims can be accomplished in a similar manner.
\end{proof}

From Lemma \ref{lemma:crashProbRanges} we now know the possible equilibrium accident probabilities in any region of parameter space. 
Using these values and Lemma \ref{lemma:EQConditions}, we can derive what a signaling equilibrium in each range must look like. 

\begin{lemma}
For any signaling game $G = (\beta, y, r)$, a unique signaling equilibrium $\xnne$ exists and takes the following form: 
\label{lemma:EQTypes}
%
%
\begin{align}
    G \in E_1 &\implies \xne = (0, (0, 0)), \label{eq:SCNCUCEqForm} \\
    G \in E_2  &\implies \xne = \left(0, \left(\frac{p^{-1}(\Pvu)}{1-Q}, 0\right)\right), \label{eq:SCNCUIEqForm} \\
    G \in E_3 &\implies \xne = (0, (y, 0)), \label{eq:SCNCUREqForm} \\
    G \in E_4 &\implies \xne = \left(p^{-1}(\Pn))-(1-Q)y, (y, 0)\right), \label{eq:SCNIUREqForm} \\
    G \in E_5 &\implies \xne = (1-y, (y, 0)), \label{eq:SCNRUREqForm} \\
    G \in E_6 &\implies \xne = \left(1-y, \left(y, \frac{p^{-1}(\Pvs) - 1 + Qy}{Q} \right)\right), \label{eq:SINRUREqForm} \\
    G \in E_7 &\implies \xne = \left(1-y, (y, y)\right). \label{eq:SRNRUREqForm} 
\end{align}
\end{lemma}

For each of the five claims, we reuse the following proof method:
\begin{enumerate}
    \item For each region, apply Lemma \ref{lemma:crashProbRanges} to obtain a condition on $P(x)$
    \item Use \eqref{eq:V2VUnsignaledBayes}, \eqref{eq:V2VSignaledBayes}, and \eqref{eq:recklesnessOrdering},  and the condition on $P(x)$ as needed to derive similar conditions on $\mathbb{P} (\A | \neg \S)$ and $\mathbb{P} (\A | \S)$
    \item Apply Lemma \ref{lemma:EQConditions} using these conditions to derive which type of equilibrium exists in that region
\end{enumerate}

\begin{proof}
We prove this in cases. 
First, assume that $G \in E_3$, i.e.
\[
    p((1 - \beta \Pvu (t(y)-f(y)) - \beta f(y))y) < \Pvu \text{ and } \Pn < p((1 - \beta \Pn (t(y)-f(y)) - \beta f(y))y).
\]
By Lemma \ref{lemma:crashProbRanges}, $\Pn < P < \Pvu$.
Then, by \eqref{eq:V2VUnsignaledBayes}, \eqref{eq:V2VSignaledBayes}, and \eqref{eq:recklesnessOrdering}, $\mathbb{P} (\A | \neg \S) < \frac{1}{1+r}$ and $\mathbb{P} (\A | \S) > \frac{1}{1+r}$.
Finally, by Lemma \ref{lemma:EQConditions}, $(0, (y, 0))$ is a signaling equilibrium and essentially unique. 

An identical method can be used to show that $(0, (0, 0))$, $(1-y, (y, 0))$, and $(1-y, (y, y))$ are essentially unique signaling equilibria if $G \in E_1$, $G \in E_5$, or $G \in E_7$, respectively. 

Now, assume that $G \in E_2$.
By Lemma \ref{lemma:crashProbRanges}, $P(\xne) = \Pvu$, and \eqref{eq:V2VUnsignaledBayes} implies that $\mathbb{P} (\A | \neg \S) = \frac{1}{1+r}$

If $\beta t(y) > 0$, then $\Pn < \Pvu$, so by \eqref{eq:V2VUnsignaledBayes}, \eqref{eq:V2VSignaledBayes}, and \eqref{eq:recklesnessOrdering}$, P(\xne) > \frac{1}{1+r}$ and $\mathbb{P} (\A | \S) > \frac{1}{1+r}$.
By Lemma \ref{lemma:EQConditions}, 
\[
    \left(0, \left(\frac{p^{-1}(\Pvu)}{1-Q}, 0\right)\right)
\]
is then an essentially unique signaling equilibrium. 

Otherwise, $\beta t(y) = 0$, so $Q(\xne) = 0$.
Note that by \eqref{eq:recklesnessOrdering}, $P(\xne) = \Pvu = \Pn > \Pvs$. 
By \eqref{eq:V2VUnsignaledBayes}, $\mathbb{P} (\A | \neg \S) = \frac{1}{1+r}$, and by \eqref{eq:V2VSignaledBayes}, $\mathbb{P} (\A | \S) > \frac{1}{1+r}$
Therefore by Lemma \ref{lemma:EQConditions}, $\xnne = p^{-1}(P(\xne)) - (1 - Q(\xne))y = p^{-1}(\frac{1}{1+r}) - y$, $\xvune = \frac{p^{-1}(P(\xne))}{1-Q(\xne)} = p^{-1} (\frac{1}{1+r})$, and $\xvsne = 0$. 
By \eqref{eq:consistency}, this gives that 
\[
    \frac{1}{1+r} = p \left(p^{-1} \left(\frac{1}{1+r} \right) - y + p^{-1} \left(\frac{1}{1+r} \right)\right),
\]
forcing $p^{-1}(\frac{1}{1+r}) = y$.
Therefore, by substitution, $(0, (y, 0))$ must be an essentially unique signaling equilibrium (note that this is a special case of the more general form given above). 
A similar technique can be used to show that signaling equilibria of the forms claimed are forced when $G \in E_4$ and $G \in E_6$. 

By Lemma \ref{lemma:crashProbRanges}, any game $G$ must satisfy at least one of the above conditions, and therefore has an essentially unique signaling equilibrium. 
\end{proof}

Finally, we are equipped to prove Theorem \ref{thm:EQCharacterization}.

\subsubsection*{Proof of Theorem~\ref{thm:EQCharacterization}}
Lemma \ref{lemma:EQTypes} demonstrated existence and essential uniqueness of a signaling equilibrium for all signaling games $G$. 
Note that each of these equilibria are consistent with the forms claimed, completing the proof. 
\hfill\QED

\section{Conclusion}
This paper has posed and analyzed a simple model of self-interested driver behavior in the presence of road hazard signals.
We have shown that warning a subset of drivers more often about traffic accidents can paradoxically lead to an increased probability of accidents occurring, relative to leaving all drivers uninformed.
For future work, it will be interesting to situate these models in the context of network routing problems or to consider more expressive signaling policies.

\begin{appendix}
\subsection{Proof of Proposition \ref{prop:consistencySol}}
Equation~\eqref{eq:consistency} defines equilibrium crash probability $P(x)$ through a rather complicated recursive relationship. 
However, we show that this relationship must always have a solution. 

\label{apx:consistencySolProof}
\subsection*{Proof of Proposition~\ref{prop:consistencySol}}
Note that by rearranging the right side of \eqref{eq:consistency}, we have that
\[
    P(x) = p(\xn + \xvu -\beta (P(x)(t(y)-f(y))+f(y))(\xvu-\xvs) ). \label{eq:consistencyAlt}
\]

Note that $P(x)$ can take on any value in the range $[p(0), p(1)]$. Therefore, consider the function $g(c) : [p(0), p(1)] \to [p(0), p(1)]$ by $g(c) := c$. 
Clearly, $g(c)$ is continuous. 
Next, consider the function $h(c) : [p(0), p(1)] \to [p(0), p(1)]$ by $h(c) := p(\xn + \xvu -\beta (c(t(y)-f(y))+f(y))(\xvu-\xvs) )$.
Since compositions of continuous functions are continuous, and $p(d)$ is continuous, $h(c)$ must also be. 
But then $g(c)$ and $h(c)$ are continuous functions bounded within the same range, so there must exist at least one $\bar{c} \in [p(0), p(1)]$ such that $g(\bar{c}) = h(\bar{c})$.
That is, \eqref{eq:consistency} must have at least one solution for $P(x)$ for all parameter combinations. 
\hfill \QED

\subsection{Counter-Examples for when $f(y) \equiv 0$}
\label{apx:noFPCounterExamples}
A potential critique of the paradoxes prevented in our paper is that false positive signals excessively deceive drivers, causing the strange equilibrium behavior. 
However, this is not the case. 
Even without any false positive signals, our two main paradoxes are still possible. 

\begin{example}
Accident probability can be increasing with $\beta$ even if $f(y) \equiv 0$. 
Let $p(d)=0.3d+0.1$, $t(y)=0.9y$, $f(y) \equiv 0$, $y=0.9$, and $r=3$.
Additionally, let $\beta_1 = 0$, $\beta_2 = 0.4004$, $G_1 = (\beta_1, y, r)$, and $G_2 = (\beta_2, y, r)$.
Then, $G_1 \in E_2$ and $G_2 \in E_2$. 
Therefore, $P(G_1) \approx 0.25$ and $P(G_2) \approx 0.3304$
Thus, increasing the quality of V2V information can increase the accident probability at equilibrium.
\end{example}

\begin{example}
Social cost can be increasing with $\beta$ even if $f(y) \equiv 0$. 
Let $p(d) = d^\frac{1}{4}$, $t(y) = 0.9y$, $f(y) \equiv 0$, $y = 0.07$, and $r = 1.001$. 
Additionally, let $\beta_1 = 0.9$, $\beta_2 = 1$, $G_1 = (\beta_1, y, r)$, and $G_2 = (\beta_2, y, r)$.
Then, $G_1 \in E_3$ and $G_2 \in E_3$.
Numerical solvers give that $S(G_1) \approx 0.4889$, while $S(G_2) \approx 0.4890$.
Thus, increasing the quality of V2V information can increase the expected cost to drivers. 
\end{example}

Recall that in general, it is possible for social cost to be increasing with $\beta$ when $G \in E_5$ (see Example~\ref{ex:SCNRURSocialCostIncreasing}). 
Interestingly, this is no longer possible if $f(y) \equiv 0$. 

\subsection{Code Availability}
All code used for this project is available at \tt{https://github.com/descon-uccs/gould-trptc-2022}. 
\end{appendix}

\vspace{-2mm}

\bibliographystyle{ieeetr}
\bibliography{ref/References,ref/library}

\end{document}